\documentclass[fleqn,twoside,twocolumn,nofootinbib,showkeys]{revtex4} 
\usepackage[nocpr]{ujp} 

\begin{document}
\title[Power Factor for Layered Thermoelectric
Materials] {POWER FACTOR FOR LAYERED\\ THERMOELECTRIC MATERIALS WITH
A CLOSED\\ FERMI SURFACE
IN A QUANTIZING MAGNETIC FIELD}%
\author{P.V. Gorskyi}
\affiliation{Ju. Fed'kovich National University of
Chernivtsi}%
\address{2 Kotsubins'kogo Str., 58012, Chernivtsi,
Ukraine}%
\email{gena.grim@gmail.com}

\autorcol{P.V.\hspace*{0.7mm}Gorskyi}

 \udk{538.9} 

\razd{\secvii}
\setcounter{page}{370}%

\begin{abstract}
The field dependence of the power factor for a layered thermoelectric
material with a closed Fermi surface in a quantizing magnetic field
and at helium temperatures has been studied in the geometry where the
temperature gradient and the magnetic field are perpendicular to the
material layers. The calculations are carried out in the constant
relaxation time approximation. In weak magnetic fields, the
layered-structure effects are shown to manifest themselves in a
phase retardation of power factor oscillations, increase of their
relative contribution, and certain reduction of the power factor
in whole. In high magnetic fields, there exists an optimal range,
where the power factor reaches its maximum, with the corresponding
value calculated for the chosen parameters of the problem in the effective
mass approximation being by 12{\%} higher than that for real layered
crystals. Despite low temperatures, the power factor maximum
obtained with those parameters in a magnetic field of 1~T has a
value characteristic of cuprate thermoelectric materials at 1000~K.
For this phenomenon to take place, it is necessary that the ratio
between the free path of charge carriers and the interlayer distance
should be equal to or larger than 30,000. However, in ultraquantum
magnetic fields, the power factor drastically decreases following
the dependence $P\propto T^{-3}B^{-6}$. The main reason for this
reduction is a squeeze of the Fermi surface along the magnetic field
in the ultraquantum limit owing to the condensation of charge
carriers on the bottom of a single filled Landau subband.
\end{abstract}


\maketitle

\section{Introduction}

Nowadays, the thermoelectric  properties of many materials are
intensively studied. The objects of researches are metals, alloys,
semiconductors \cite{1,2}, fullerenes \cite{3}, composites \cite{4},
including biomorphic ones \cite{5}, {\it etc}. A promising
thermoelectric material is graphene \cite{6}. The specific
photothermoelectric effect in graphene, which was considered earlier
as of exclusively the photovoltaic origin, allows it to be
considered as a material for solar cells with a high efficiency.

The theory of thermoelectric  properties for various substances,
including nanosystems \cite{7,8}, is intensively developed. For
instance, one of the first works devoted to the theory of transverse
thermoelectric coefficients of metals in a quantizing magnetic field
has been published by A.M.~Kosevich and
\mbox{V.V.~Andreev~\cite{9}.}

A large number of substances under investigation~-- e.g., the
semiconductor A$^{\mathrm{III}}$B$^{\mathrm{VI}}$C$^{\mathrm{VII}}$
systems, intercalated compounds of graphite, synthetic metals on the
basis of organic compounds, graphene, and others belong to the
layered materials according to their crystal structure. At the same
time, in the overwhelming majority of theoretical works dealing with
the behavior of such layered systems in quantizing magnetic fields,
only the transverse galvano-magnetic effects are studied as a rule.
The thermal conductivity of graphene in a magnetic field, including
the quantizing one, its electric dc and ac conductivities, and the
quantum Hall effect were considered in works \cite{10,11}. In those
works, the Fermi surface (FS) of graphene was assumed to be open,
i.e., it occupies the whole one-dimensional Brillouin zone and is
connected, if being periodically continued; in other words, it looks
like a continuous corrugated cylinder. The cited works were mainly
devoted to the research of the charge carrier behavior in a layer
plane. Therefore, the concept of \textquotedblleft massless
neutrino\textquotedblright\ with the linear dependence of the energy
on the quasimomentum in the layer plane turned out to be an
effective tool for the description of the band structure in
graphene. In this case, the energy levels in a quantizing magnetic
field were determined in the framework of the effective Hamiltonian
method.

On the other hand, in the previous works,  the author showed that,
if the behavior of charge carriers in the direction perpendicular to
the layers is studied, the layered-structure effects can be well
pronounced in the case of a closed FS as well \cite{12} or if the
topological transition from the open FS to a closed one takes place
\cite{13}.

It should be noticed that the power factor,  which equals the
product of the squared thermoelectric coefficient and the
conductivity, is rather an indicative integrated characteristic for
the system of free charge carriers in a material \cite{14}. This is
of importance for the estimation of thermoelectric properties of and
applications. Therefore, this work was aimed at calculating and
researching the dependence of the power factor for a layered
thermoelectric material with a closed FS on the induction of a
quantizing magnetic field.

\section{Calculation of the Power Factor for~a~Layered Crystal and
Discussion of~the~Results Obtained}

While calculating the power  factor for a layered crystal in a
quantizing magnetic field directed perpendicularly to the layer
planes, the following dispersion law for the charge carriers will be
used:
\begin{equation}
\varepsilon\left(  {n,x}\right)  =\mu^{\ast}B\left(  {2n+1}\right)
+W\left(
x\right)\!,\label{eq1}%
\end{equation}
where $\mu^{\ast}=\mu_{\rm B}{m_{0}}/m^{\ast}$, $\mu_{\rm B}$ is the
Bohr magneton, ${m_{0}}$ the free electron mass, $m^{\ast}$ the
effective electron mass in a layer plane, $B$ the magnetic field
induction, $n$ the Landau level number, $W\left(  x\right)$
the dispersion law for charge carriers along the superlattice axis,
$x=ak_{z}$, $k_{z}$ is the component of the quasimomentum along the
superlattice axis, and $a$ the distance between translationally
equivalent layers.

In order to study the influence  of layered-structure effects on the
power factor, the latter will be calculated in two cases. These are
the strong coupling approximation,
\begin{equation}
W\left(  x\right)  =\Delta\left(  {1-\cos x}\right)\!,\label{eq2}%
\end{equation}
where $\Delta$ is the miniband  halfwidth that governs the motion of
electrons between the layers, and the case where only the term
quadratic in $x$ is retained in the series expansion of formula
(\ref{eq2}), which corresponds to the effective mass approximation.
In both cases, the dependence of chemical potential in the electron
gas on the magnetic field induction is taken into account. To
simplify calculations, the relaxation time for charge carriers will
be considered constant. With the same purpose in view, the influence
of Dingle factor on longitudinal conductivity oscillations
(Shubnikov--de Haas oscillations) and oscillations of the
longitudinal thermoelectric coefficient $\alpha_{zz}$ (the latter is
defined, to within the sign accuracy, as the proportionality
coefficient between the electrochemical potential gradient in the
system of charge carriers and the temperature gradient, provided
that the current equals zero). It should be noticed, however, that
the indicated factor can play a substantial role at the electron
scattering by impurities and defects in the crystal lattice
\cite{15}.

If we proceed from the  Boltzmann kinetic equation and act in the
same way as when deriving the formula for the longitudinal
electroconductivity, the following general formula for the
thermoelectric coefficient $\alpha_{zz}$ can be obtained:
\[
\alpha _{zz} =-\left[ {\frac{\partial }{\partial
T}\sum\limits_\gamma {\tau _\gamma g_\gamma v_{z\gamma }^2 f^0\left(
{\varepsilon _\gamma } \right)} } \right]\times
\]\vspace*{-7mm}
\begin{equation}
\label{eq3} \times\left[ {e\frac{\partial }{\partial \zeta
}\sum\limits_\gamma {\tau _\gamma g_\gamma v_{z\gamma }^2 f^0\left(
{\varepsilon _\gamma } \right)} } \right]^{-1}\!.
\end{equation}
Here, $T$ is the absolute temperature,  $e$ the elementary charge,
$\zeta$ the chemical potential of the system of charge carriers (we
consider them to be electrons), $\gamma\equiv\left({n,x}\right)$ a
set of quantum numbers that characterize the electron energy,
$\tau_{\gamma}$ the relaxation time for charge carriers,
$v_{z\gamma}$ the longitudinal electron velocity, $g_{\gamma }$ the
statistical weight of a Landau level per unit volume of the crystal,
and $f^{0}\left(  {\varepsilon_{\gamma}}\right)$ the Fermi--Dirac
distribution function. For the dispersion law (\ref{eq1}) and
provided that $\tau_{\gamma }\equiv\tau\left(  x\right)$, formula
(\ref{eq3}) becomes
\begin{equation}
\alpha_{zz}=\frac{\pi k}{e}\frac{A}{B+C}.\label{eq4}%
\end{equation}
The quantities $A$, $B$, and $C$ are
\[
A\!=\!\sum\limits_{l=1}^\infty \left( {-1} \right)^{l-1}f_l^{\rm th}
\!\!\! \!\!\!\!\int  \limits_{W\left( x \right)\le \zeta
}\!\!\!\!\!\!\! \tau \left( x \right){W}'\left( x \right)^2\times
\]
\begin{equation}
\label{eq5} \times\sin \left[ {\pi l\frac{\zeta -W\left( x
\right)}{\mu ^\ast B}} \right]dx,
\end{equation}\vspace*{-5mm}
\begin{equation}
\label{eq6} B=0.5\int\limits_{W\left( x \right)\le \zeta } {\tau
\left( x \right){W}'\left( x \right)^2dx},
\end{equation}\vspace*{-5mm}
\[
C\!=\!\sum\limits_{l=1}^\infty \left( {-1} \right)^lf_l^\sigma
\!\!\! \!\!\!\int\limits_{W\left( x \right)\le \zeta }\!\!\! \!\!\!
\tau \left( x \right){W}'\left( x \right)^2\times
\]\vspace*{-5mm}
\begin{equation}
\label{eq7} \times\cos \left[ {\pi l\frac{\zeta -W\left( x
\right)}{\mu ^\ast B}} \right]dx,
\end{equation}
where
\[
 f_l^{\rm th} =\left[ {{\mathrm{sh}}\left( {{\pi ^2lkT}
\mathord{\left/ {\vphantom {{\pi ^2lkT} {\mu ^\ast B}}} \right.
\kern-\nulldelimiterspace} {\mu ^\ast B}} \right)}
\right]^{-1}\times
\]\vspace*{-7mm}
\begin{equation}
\label{eq8} \times\left[ {1-\left( {{\pi ^2lkT} \mathord{\left/
{\vphantom {{\pi ^2lkT} {\mu ^\ast B}}} \right.
\kern-\nulldelimiterspace} {\mu ^\ast B}}
\right){\mathrm{cth}}\left( {{\pi ^2lkT} \mathord{\left/ {\vphantom
{{\pi ^2lkT} {\mu ^\ast B}}} \right. \kern-\nulldelimiterspace} {\mu
^\ast B}} \right)} \right]\!,
\end{equation}\vspace*{-7mm}
\begin{equation}\label{9}
f_l^\sigma =\left( {{\pi ^2lkT} \mathord{\left/ {\vphantom {{\pi
^2lkT} {\mu ^\ast B}}} \right. \kern-\nulldelimiterspace} {\mu ^\ast
B}} \right)\left[ {{\mathrm{sh}}\left( {{\pi ^2lkT} \mathord{\left/
{\vphantom {{\pi ^2lkT} {\mu ^\ast B}}} \right.
\kern-\nulldelimiterspace} {\mu ^\ast B}} \right)} \right]^{-1}\!.
\end{equation}
In accordance with the results of work \cite{12}, the total
longitudinal electric conductivity of a layered crystal can be
written down in the form
\begin{equation}
\sigma_{zz}=\frac{32\pi^{2}e^{2}m^{\ast}a}{h^{4}}\left(
{B+C}\right)\!
.\label{eq9}%
\end{equation}
Hence, the power factor equals
\begin{equation}
P=\frac{32\pi^{4}k^{2}m^{\ast}aA^{2}}{h^{4}\left(  {B+C}\right)
}.\label{eq10}
\end{equation}
In the case of the dispersion law (\ref{eq2}) and $\tau\left(  x\right)
\equiv\tau_{0}$, it is easy to pass to the dimensionless coefficients
$A$,
$B$, and $C$ and write down the power factor as
\begin{equation}
P=\frac{16\pi^{4}k^{2}m^{\ast}a\Delta^{2}\tau_{0}A^{2}}{h^{4}\left(
{B+C}\right)  }.\label{eq11}%
\end{equation}
Now, the coefficients $A$, $B$, and $C$ are as follows:
\[
A=\sum\limits_{l=1}^\infty ( {-1} )^{l-1}f_l^{\mathrm{th}}
\biggl\{\! \sin \left(\! {\pi l\frac{\zeta -\Delta }{\mu ^\ast B}}
\!\right)\!\biggl[ \!\left( {C_0 -C_2 } \right)\times
\]\vspace*{-5mm}
\[
\times J_0 \left(\! \frac{\pi l\Delta }{\mu ^\ast B}
\!\right)+\sum\limits_{r=1}^\infty {\left( {-1} \right)^r\left(
2C_{2r} -C_{2r+2} -C_{2r-2}  \right)}\times
\]\vspace*{-5mm}
\[
\times J_{2r} \left(\! \frac{\pi l\Delta }{\mu ^\ast B}\! \right)\!
\biggr] +\cos \left(\! \pi l\frac{\zeta -\Delta }{\mu ^\ast B}
\!\right) \sum \limits_{r=0}^\infty {\left( {-1} \right)^r }\times
\]\vspace*{-5mm}
\begin{equation}
\label{eq12}\times\left( 2C_{2r+1} -C_{2r+3} -C_{\left| {2r-1}
\right|}  \right )J_{2r+1} \left(\! \frac{\pi l\Delta }{\mu ^\ast B}
\!\right)\! \biggr\}\!,
\end{equation}
\begin{equation}
B=0.5\left(  {C_{0}-C_{2}}\right) \!.\label{eq13}%
\end{equation}\vspace*{-7mm}
\[
C=\sum\limits_{l=1}^\infty \left( {-1} \right)^lf_l^\sigma \biggl\{
\cos \left(\! \pi l\frac{\zeta -\Delta }{\mu ^\ast B}
\!\right)\!\biggl[ \left( {C_0 -C_2 } \right)\times
\]\vspace*{-5mm}
\[
\times J_0 \left(\! {\frac{\pi l\Delta }{\mu ^\ast B}} \!\right)+
\sum\limits_{r=1}^\infty {\left( {-1} \right)^r\left( {2C_{2r}
-C_{2r+2} -C_{2r-2} } \right)}\times
\]\vspace*{-5mm}
\[
\times J_{2r} \left( \!\frac{\pi l\Delta }{\mu ^\ast B}\! \right)\!
\biggr]-\sin \left(\! \pi l\frac{\zeta -\Delta }{\mu ^\ast B}
\!\right)\sum\limits_{r=0}^\infty \left( {-1} \right)^r\times
\]\vspace*{-5mm}
\begin{equation}
\label{eq14} \times( 2C_{2r+1} -C_{2r+3} -C_{\left| {2r-1} \right|}
) J_{2r+1} \left(\! {\frac{\pi l\Delta }{\mu ^\ast B}} \!\right)\!
\biggr\}\!.
\end{equation}
In formulas (\ref{eq12})--(\ref{eq14}), $J_{n}\left(  x\right)$
are the Bessel functions of the first kind of a real argument. The
modulating coefficients $C_{m}$ are defined as follows:
\begin{equation}
\label{eq15} C_0 =\arccos \left( \!{1-\frac{\zeta -\mu ^\ast
B}{\Delta }} \!\right)\!,
\end{equation}\vspace*{-5mm}
\begin{equation}
C_m =\frac{\sin mC_0 }{m}$ for $m\ne 0.
\end{equation}
In the approximation concerned, formulas (\ref{eq11})--(17)
completely determine the temperature and field dependences of the power
factor for a layered crystal, with the chemical potential of
a gas of charge carriers in formulas (\ref{eq15}) and (17) being
considered to depend on the magnetic field. Owing to the FS
squeezing in a crystal in the quantizing magnetic field, the
quantity $\mu^{\ast}B$, i.e. the minimum electron energy, is
subtracted from the chemical potential.

In the effective mass approximation, the dimensionless coefficients $A$,
$B$,
and $C$ look like
\[
A=\sum\limits_{l=1}^\infty ( -1)^{l-1}f_l^{\rm th} \Biggl\{
\frac{\mu ^\ast B}{\pi l\Delta }\sqrt {\frac{2\left( {\zeta -\mu
^\ast B} \right)}{\Delta }} -
\]\vspace*{-5mm}
\[
-\frac{1}{\pi }\left( \!\frac{\mu ^\ast B}{l\Delta }
\!\right)^{\!3/2}\Biggl[ \cos \left(\! {\frac{\pi l\zeta }{\mu ^\ast
B}} \!\right) C\left(\! \sqrt {2l\left(\! {\frac{\zeta }{\mu ^\ast
B}-1} \!\right)}\!  \right)+
\]\vspace*{-5mm}
\begin{equation}
\label{eq16} +\sin \left(\! {\frac{\pi l\zeta }{\mu ^\ast B}}
\!\right)S\left(\! {\sqrt {2l\left( \!{\frac{\zeta }{\mu ^\ast B}-1}
\!\right)} } \!\right)\! \Biggr] \!\Biggr\}\!,
\end{equation}\vspace*{-5mm}
\begin{equation}
B=\frac{1}{6}\left[\!  {2\left(\!
{\frac{\zeta-\mu^{\ast}B}{\Delta}}\!\right)\! }\right]
^{\!3\mathord{\left/ {\vphantom {3 2}} \right.
\kern-\nulldelimiterspace}2}\!,
\label{eq17}%
\end{equation}\vspace*{-5mm}
\[
C= \sum\limits_{l=1}^\infty ( -1)^lf_l^\sigma \frac{1}{\pi }\left(\!
\frac{\mu ^\ast B}{l\Delta } \!\right)^{\!3/2}\!\Biggl[ \sin
\left(\! {\frac{\pi l\zeta }{\mu ^\ast B}}\! \right)\times
\]
\[
\times C\left( \!\sqrt {2l\left(\! \frac{\zeta }{\mu ^\ast B}-1
\!\right)}\!\right)-\cos \left(\! \frac{\pi l\zeta }{\mu ^\ast B}
\!\right)\times
\]\vspace*{-5mm}
\begin{equation}
\label{eq18}\times S\left(\! \sqrt {2l\left(\! \frac{\zeta }{\mu
^\ast B}-1 \!\right)} \!\right)\!\Biggr]\!.
\end{equation}
Here, $C\left(  x\right)  $ and $S\left(  x\right)  $ are the Fresnel
cosine
and sine integrals, respectively.

For further calculations,  the dependence of chemical potential for
the gas of charge carriers on the quantizing magnetic field is
required. In work \cite{12}, this dependence was already presented,
but it is useful to recall it. The equation that determines the
chemical potential of the electron gas in a quantizing magnetic
field at low temperatures has the following general form:
\[
n_0 =\frac{4m^\ast }{ah^2}\!\!\int\limits_{W\left( x \right)\le
\zeta } \!\!\!\!{\left[ {\zeta -W\left( x \right)} \right]dx}
+\frac{8\pi m^\ast kT}{ah^2}\times
\]\vspace*{-7mm}
\begin{equation}
\label{eq19} \times \sum\limits_{l=1}^\infty {\frac{\left( {-1}
\right)^l}{{\mathrm{sh}}\left( {{\pi ^2lkT} \mathord{\left/
{\vphantom {{\pi ^2lkT} {\mu ^\ast B}}} \right.
\kern-\nulldelimiterspace} {\mu ^\ast B}} \right)}}\!\!
\int\limits_{W\left( x \right)\le \zeta }\!\!\!\! {\sin \left(\!
{\pi l\frac{\zeta -W\left( x \right)}{\mu ^\ast B}} \!\right)dx},
\end{equation}
where $n_{0}$ is the bulk  concentration of charge carriers in the
crystal. For the dispersion law (\ref{eq2}) and in the case of a
closed FS, Eq.~(\ref{eq19}) reads
\[
n_0 =\frac{4m^\ast \Delta }{ah^2}\left[ {\left( {\gamma -1}
\right)C_0 +\sqrt {2\gamma -\gamma ^2} } \right]+
\]\vspace*{-7mm}
\[
+\frac{8m^\ast \pi kT}{ah^2}\sum\limits_{l=1}^\infty {\frac{\left(
{-1} \right)^l}{{\mathrm{sh}}\left( {{\pi ^2lkT} \mathord{\left/
{\vphantom {{\pi ^2lkT} {\mu ^\ast B}}} \right.
\kern-\nulldelimiterspace} {\mu ^\ast B}} \right)}} \Bigg\{\! {\sin
\left( \!{\pi l\frac{\zeta -\Delta }{\mu ^\ast B}} \!\right)\times }
\]\vspace*{-7mm}
\[
\times \left[ {C_0 J_0 \left(\! {\frac{\pi l\Delta }{\mu ^\ast B}}
\!\right)+2\sum\limits_{r=1}^\infty {\left( {-1} \right)^rC_{2r} }
J_{2r} \left(\! {\frac{\pi l\Delta }{\mu ^\ast B}} \!\right)}\!
\right]+
\]\vspace*{-7mm}
\begin{equation}
\label{eq20} {+{ 2\cos}\left(\! {\pi l\frac{\zeta -\Delta }{\mu
^\ast B}} \!\right)\sum\limits_{r=0}^\infty {\left( {-1}
\right)^rC_{2r+1} J_{2r+1} \left(\! {\frac{\pi l\Delta }{\mu ^\ast
B}}\! \right)} }\! \Bigg\}\!.
\end{equation}
The modulating coefficients $C_{m}$ in this equation should be taken at
$B=0$,
with $\gamma=\zeta/\Delta$ at that.

\begin{figure}
\vskip1mm
\includegraphics[width=4.5cm]{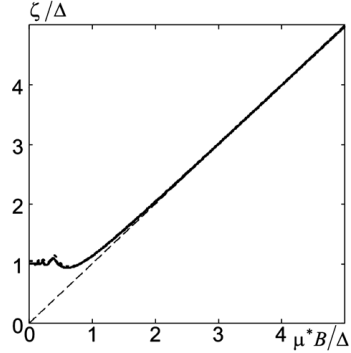}
\vskip-3mm\caption{Field dependence of the chemical potential at
$\gamma_{0}=$ $=1$ and ${kT/\Delta=0.03}$ for a layered crystal
(solid curve) and in the effective mass approximation (dashed
curve). Thin dashed line corresponds to the limiting case
$\zeta=\mu^{\ast}B$  }
\end{figure}

In the case of an open FS, i.e. at $\gamma-{\mu^{\ast}B}/\Delta\geq2$,
we have to put $C_{0}=\pi$ in formulas (\ref{eq12})--(17) and
(\ref{eq20}), and, in addition, consider the radical in formula
(\ref{eq20}) equal to zero.

In the effective mass approximation,  Eq.~(\ref{eq20}) takes the form%
\[
n_0 =\frac{8m^\ast \zeta }{3ah^2}\sqrt {\frac{2\zeta }{\Delta }}
+\frac{8\pi m^\ast kT}{ah^2}\sqrt {\frac{\mu ^\ast B}{\Delta }}
\times
\]\vspace*{-5mm}
\[
\times \sum\limits_{l=1}^\infty {\frac{\left( {-1} \right)^l}{l^{1
\mathord{\left/ {\vphantom {1 2}} \right. \kern-\nulldelimiterspace}
2}{\rm sh}\left(\! {{\pi ^2lkT} \mathord{\left/ {\vphantom {{\pi
^2lkT} {\mu ^\ast B}}} \right. \kern-\nulldelimiterspace} {\mu ^\ast
B}} \right)}} \left[ {\sin \left(\! {\frac{\pi l\zeta }{\mu ^\ast
B}} \!\right)C\left(\! {\sqrt {\frac{2l\zeta }{\mu ^\ast B}} }\!
\right)} \right.\!\!+
\]\vspace*{-5mm}
\begin{equation}
\label{eq21} \left. {+\cos \left(\! {\frac{\pi l\zeta }{\mu ^\ast
B}} ~\right)S\left(\! {\sqrt {\frac{2l\zeta }{\mu ^\ast B}} }
\!\right)} \!\right]\!.
\end{equation}
The concentrations of charge carriers in both examined cases will be
considered identical and determined by the formula
\begin{equation}
\label{eq22} n_0 =\frac{4m^\ast \Delta }{ah^2}\!\left[\! {\left(
{\gamma _0 \!-\!1} \right)\arccos \left( {1\!-\!\gamma _0 }
\right)\!+\!\sqrt {2\gamma _0\! \!-\gamma _0^2 } } \right]\!\!,\!
\end{equation}
where $\gamma_{0}={\zeta_{0}}/\Delta$, and ${\zeta_{0}}$ is the
Fermi energy of the crystal electron gas with the dispersion law
(\ref{eq2}) at the zero temperature and in the absence of a
\mbox{magnetic field.}

The solutions of Eqs.~(\ref{eq20}) and (\ref{eq21}) can be found
numerically, The corresponding results obtained for the layered
crystal with a closed FS is presented in Fig.~1 in the range
$0<\mu^{\ast}B/\Delta<5$. If the magnetic field grows, both curves
approach each other, because Eq.~(\ref{eq20}) has the following
asymptotic solution in strong quantizing magnetic fields in the case
of a closed FS \cite{12}:
\begin{equation}
\zeta\left(  B\right)  =\mu^{\ast}B+\Delta\left[  {1-\cos\left(\!
{\frac{f\left(  {\gamma_{0}}\right)
\Delta}{2\mu^{\ast}B}}\!\right)\! }\right]\!
.\label{eq23}%
\end{equation}
In the case of a closed FS,
\begin{equation}
f\left(  {\gamma_{0}}\right)  =\left(  {\gamma_{0}-1}\right)
\arccos\left(
{1-\gamma_{0}}\right)  +\sqrt{2\gamma_{0}-\gamma_{0}^{2}},\label{eq24}%
\end{equation}
whence one can see that the single  filled Landau subband gets
narrower in the ultraquantum limit, and this narrowing has to be
taken into account while calculating the power factor.

In the effective mass approximation, formula (\ref{eq23}) acquires the
form
\begin{equation}
\zeta\left(  B\right)  =\mu^{\ast}B+\frac{\Delta^{3}f^{2}\left(
{\gamma_{0}%
}\right)  }{8\left(  {\mu^{\ast}B}\right)  ^{2}}. \label{eq25}%
\end{equation}
Note that all the formulas presented above for the longitudinal
thermoelectric coefficient and the power factor were obtained under
the condition that the interaction-induced broadening of energy
levels is small in comparison with the distance between Landau
levels. Only in this case, the broadening of energy levels can be
directly connected with the relaxation time, and the shift of energy
levels associated with the scattering can be neglected. Then, the
approach based on the Boltzmann equation is completely equivalent to
that based on the Kubo formalism.

For the calculation of the power factor, we also need an expression for the
relaxation time $\tau_{0}$. For any $\gamma_{0}$, this quantity can be
determined in the same way as in works \cite{12,13}, namely, by the
formula
\begin{equation}
\tau_{0}=\frac{lm_{es}^{\ast}}{hk_{0}}.\label{eq26}%
\end{equation}
Here, $l$ is the mean free path of  charge carriers, which is
determined by the scattering at impurities; $k_{0}$ the radius of an
equivalent sphere, by which the real FS is substituted for the
scattering to be considered as isotropic; and $m_{es}^{\ast}$ the
equivalent mass of charge carriers at this sphere. The two latter
parameters are so defined that the concentration of charge carriers
and the Fermi energy coincide with the corresponding parameters
determined either for the real layered crystal or in the effective
mass approximation. Therefore, in both cases,
\begin{equation}
k_{0}=\sqrt[3]{\frac{12\pi^{2}m^{\ast}\Delta f\left(
{\gamma_{0}}\right)
}{ah^{2}}}.\label{eq27}%
\end{equation}
At the same time, for a real layered crystal, the parameter
$m_{es}^{\ast}$
is determined by the formula
\begin{equation}
m_{es}^{\ast}=\frac{h^{2}k_{0}^{2}}{8\pi^{2}\gamma_{0}\Delta},\label{eq28}%
\end{equation}
and, in the effective mass approximation, by the
\mbox{formula}\vspace*{-2mm}
\begin{equation}
m_{es}^{\ast}=\frac{h^{2}k_{0}^{2}}{8\pi^{2}\gamma_{0em}\Delta}.\label{eq29}%
\end{equation}
Therefore, we obtain the following ultimate expression for the power
factor of
the crystal:
\begin{equation}
P=P_{0}\frac{A^{2}}{B+C},\label{eq30}%
\end{equation}
where the parameter $P_{0}$ is determined by the formula
\begin{equation}
P_{0}=\frac{2\pi^{3}k^{2}m^{\ast}a^{2}\Delta}{h^{3}\gamma_{0}}\sqrt[3]%
{\frac{12\pi^{2}m^{\ast}\Delta f\left(  {\gamma_{0}}\right)  }{ah^{2}}%
}N\label{eq31}%
\end{equation}
for a real layered crystal and by the formula
\begin{equation}
P_{0}=\frac{2\pi^{3}k^{2}m^{\ast}a^{2}\Delta}{h^{3}\gamma_{0em}}\sqrt[3]
{\frac{12\pi^{2}m^{\ast}\Delta f\left(  {\gamma_{0}}\right)  }{ah^{2}}%
}N\label{eq32}%
\end{equation}
in the effective mass approximation. The coefficients $A$, $B$, and $C$
in
formula (\ref{eq30}) are determined by formulas
(\ref{eq12})--(\ref{eq14}) for
the real layered crystal and by formulas (\ref{eq16})--(\ref{eq18}) in
the
effective mass approximation. In addition, $N\equiv l/a$.

The calculation results for the  field dependence of the power
factor obtained for a layered crystal in various ranges of the
quantizing magnetic field induction are depicted in Figs.~2 and 3.
From Fig.~2, one can see that, in the case of a real layered
crystal, the oscillations of the power factor are more pronounced,
but the corresponding absolute values are smaller in comparison with
those in the effective mass approximation. Similarly to the case of
longitudinal conductivity, this fact can be explained, because, on
the one hand, any confinement imposed on the free motion of charge
carriers reduces the conductivity and, on the other hand, the
dependence of the cross-section area of the FS by the plane
perpendicular to the magnetic field direction on the longitudinal
quasimomentum is weaker. Moreover, in the case of a real layered
crystal, a certain phase delay of oscillations takes place. This
happens, because the Fermi energy for a real layered crystal is
slightly lower than that in the effective mass approximation,
provided the same concentration of charge carriers. The same factor
is also responsible for the increase of the relative contribution of
oscillations to the power factor.

In the conventional quasiclassical approximation, the power factor
equals
\[
P=P_0 \frac{3\left( {\mu ^\ast B} \right)^3}{2^{3/2}\pi ^2\Delta ^{3
/ 2}\zeta ^{3/2}}\times
\]\vspace*{-7mm}
\begin{equation}
\label{eq33}\times \left[ {\sum\limits_{l=1}^\infty {\left( {-1}
\right)^{l-1}l^{-3/2}f_l^{\rm th} \cos \left(\! {\frac{\pi l\zeta
}{\mu ^\ast B}-\frac{\pi }{4}} \!\right)} }\! \right]^2\!.
\end{equation}
This formula also makes it evident that, if  the Fermi energy is
constant, it is impossible to distinguish the influence of
layered-structure effects on the power factor. However, since the
Fermi energy for a real layered crystal is slightly lower than that
in the effective mass approximation at a constant concentration of
charge carriers, the power factor oscillations have to be more
pronounced in the former case, which really takes place. At the same
time, Fig.~2 testifies that the power factor is a little larger in
the effective mass approximation than that in the case of real
layered crystal, although formula (\ref{eq33}), with regard for
formulas (\ref{eq31}) and (\ref{eq32}), brings about the opposite
conclusion. Such an apparent contradiction arises because the value
of power factor is also affected by the monotonous component of the
thermoelectric coefficient, which cannot be taken into account in
the framework of the traditional approach, for which the specific
shape and the size of a FS are insignificant.

From Fig.~3, one can see that, in strong quantizing magnetic fields,
the power factor oscillations are also more pronounced in the case
of real layered crystal than those in the effective mass
approximation. However, after having passed the last oscillation
minimum, the power factor starts to grow considerably and reaches a
maximum approximately at the point of the last oscillation peak for
the chemical potential. The presence of this maximum can be
explained by the following physical reasoning. On the one hand, it
is clear that, in accordance with the general thermodynamic
relations~-- with which formula (\ref{eq33}), as well as formulas
(\ref{eq12}) and (\ref{eq16}) and Figs.~2 and 3, are in full
agreement~-- the power factor has tend to zero at weak enough
magnetic fields and low enough temperatures. On the other hand, it
should also tend to zero in the ultraquantum limit as a result of
the FS squeezing along the magnetic field direction. However, this
factor cannot be equal to zero identically. Hence, it has to possess
a maximum at a certain induction of the magnetic field. In addition,
the figure demonstrates that the power factor maximum is larger and
more pronounced in the effective mass approximation than that in the
case of real layered crystal. It is explained by a higher electron
density of states and its more drastic dependence on the energy for
a real layered crystal. However, for the same reason, the power
factor after passing the maximum falls down more slowly in the case
of real layered crystal than that in the effective mass
approximation.

\begin{figure}
\vskip1mm
\includegraphics[width=4.5cm]{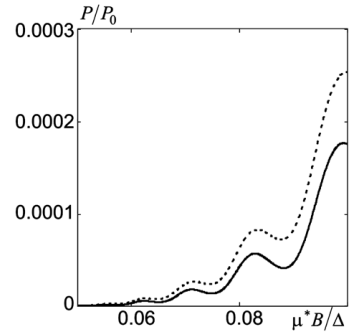}
\vskip-3mm\caption{Field dependences of the power factor at
$\gamma_{0}=1$ and $kT/\Delta =0.03$ calculated for a real layered
crystal (solid curve) and in the effective mass approximation
(dotted curve) in the interval $0.05\leq
{\mu^{\ast}B}/\Delta\leq0.1$  }\vskip3mm
\end{figure}

\begin{figure}
\includegraphics[width=4.5cm]{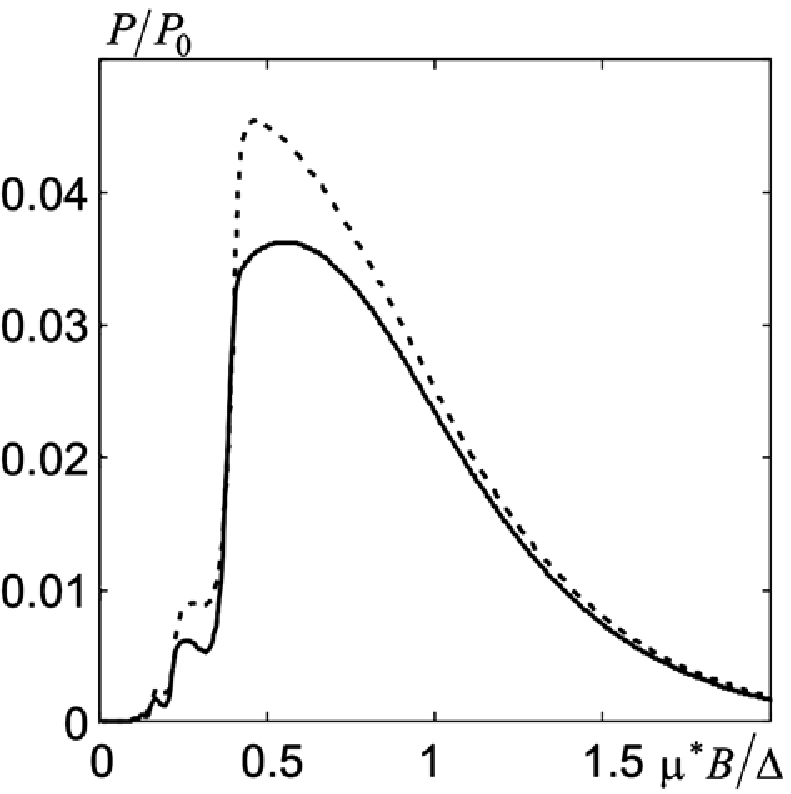}
\vskip-3mm\caption{The same as in Fig.~2, but in the interval
\mbox{$0\leq{\mu^{\ast}B}%
/\Delta\leq2$}
  }
\end{figure}

The results of numerical calculations show that,  for instance, at
$\Delta=0.01~\mathrm{eV}$, $m^{\ast}=0.01m_{0}$, $kT/\Delta=$
$=0.03$, $a=1\mathrm{~nm}$, and $\gamma_{0}=1$, the power factor in
the maximum equals $3.350\times10^{-9}N~\mathrm{W}/(\mathrm{m\cdot
K}^{2})$ for a real layered crystal and
$3.758\times10^{-9}N~\mathrm{W}/(\mathrm{m\cdot K}^{2})$ in the
effective mass approximation. On the other hand, promising are those
thermoelectric materials, for which the power factor equals, e.g.,
$1.04\times$ $\times10^{-4}~\mathrm{W}/(\mathrm{m\cdot K}^{2})$~--
to tell the truth, this value was obtained at 1000$~\mathrm{K}$
\cite{14}. Under our conditions, such a value of power factor
corresponds to $N=(  {2.767\div 3.104}) \times10^{4}$, i.e. the mean
free path of charge carriers is $l={28\div31}$~$\mu\mathrm{m}$. For
such a ratio between the mean free path and the distance between the
neighbor layers, putting the Dingle factor identically equal to 1
within the whole range of examined magnetic fields is completely
eligible. Those data can be regarded as a specific variant of
\textquotedblleft technological requirements\textquotedblright\ to
the thermoelectric material. Provided the same parameters, the
magnetic field induction, at which the power factor reaches its
maximum, amounts to approximately 1.018~T.

At last, let us derive  the asymptotic law for the power factor
recession in the ultraquantum limit. For this purpose, expression
(\ref{eq25}) should be substituted for $\zeta$ in expressions
(\ref{eq5})--(\ref{eq7}) for the coefficients $A$, $B$, and $C$. As
a result, $\left(  {-1}\right)  ^{l}$ becomes compensated owing to
the trigonometric multipliers. Then, the cosines and the sines are
put equal to 1 and to their arguments, respectively. The expression
$f_{l}^{th}$ at ${\mu^{\ast}B/kT\gg1}$ looks like
\begin{equation}
\label{eq34} f_l^{\mathrm{th}} =\frac{\left( {{\pi ^2lkT}
\mathord{\left/ {\vphantom {{\pi ^2lkT} {\mu ^\ast B}}} \right.
\kern-\nulldelimiterspace} {\mu ^\ast B}}
\right)^2}{2{\mathrm{sh}}\left( {{\pi ^2lkT} \mathord{\left/
{\vphantom {{\pi ^2lkT} {\mu ^\ast B}}} \right.
\kern-\nulldelimiterspace} {\mu ^\ast B}} \right)}.
\end{equation}
While summing up the series over $l$, let us take into account that the
numerical analysis testifies to the validity of the following
relations at small $\delta$:
\begin{equation}
\sum\limits_{l=1}^{\infty}{\frac{\delta l}{\sinh\left(  {\delta
l}\right)
}=\frac{2.467}{\delta}},\label{eq35}%
\end{equation}\vspace*{-5mm}
\begin{equation}
\sum\limits_{l=1}^{\infty}{\frac{\delta l^{3}}{\sinh\left(  {\delta
l}\right)
}=\frac{12.176}{\delta^{3}}}.\label{eq36}%
\end{equation}
The integrands in formulas (\ref{eq5})--(\ref{eq7}) are expanded in
series in $x$ to an accuracy of the leading terms and integrated
within the limits from 0 to $\frac{f\left(  {\gamma_{0}}\right)
\Delta}{2\mu^{\ast}B}$. After carrying out all transformations and
combining numerical multipliers, the following asymptotic expression
is obtained for the power factor of a real layered crystal near the
ultraquantum limit:
\[
P=1.929\times 10^{-2}k^2f^7\left( {\gamma _0 } \right)\frac{m^\ast
a^2\Delta ^{10}}{h^3\left( {\mu ^\ast B} \right)^6\left( {kT}
\right)^3\gamma _0 }\times
\]\vspace*{-5mm}
\begin{equation}
\times \sqrt[3]{\frac{m^\ast \Delta f\left( {\gamma _0 }
\right)}{ah^2}}N. \label{39}
\end{equation}
In the effective mass approximation, the power factor of the crystal
equals
\[
P=1.929\times 10^{-2}k^2f^7\left( {\gamma _0 } \right)\frac{m^\ast
a^2\Delta ^{10}}{h^3\left( {\mu ^\ast B} \right)^6\left( {kT}
\right)^3\gamma _{0em} }\times
\]\vspace*{-7mm}
\begin{equation}
\times \sqrt[3]{\frac{m^\ast \Delta f\left( {\gamma _0 }
\right)}{ah^2}}N. \label{40}
\end{equation}
Hence, for the parameters given above and at $B=$ $=60~\mathrm{T}$,
we obtain $P=3.780\times10^{-11}~\mathrm{W}/(\mathrm{m\cdot K}^{2})$
for a real layered crystal and \mbox{$P=3.635\times$}\linebreak
\mbox{$\times10^{-11}~\mathrm{W}/(\mathrm{m\cdot K}^{2})$} in the
effective mass approximation. Thus, the power factor decreases by
six orders of magnitude in comparison with its maximum value.
Therefore, although the asymptotic law \mbox{$P\propto
B^{-6}T^{-3}$} is a little hard to understand, because the power
factor does not vanish at $T=0$, which contradicts, at first sight,
the general thermodynamic principles, this law does not lead to
incorrect physical consequences at real low temperatures.

It should be noted that if the  same approach as that used in the
calculation of the longitudinal conductivity in works \cite{12,16}
be applied while calculating the power factor at low temperatures
near the ultraquantum limit, the thermoelectric coefficient, as well
as the power factor, would be identically equal to zero, because the
sum under the sign of derivative in the numerator of formula
(\ref{eq3}) does not depend on the temperature in the framework of
this approach. Therefore, this approach all the same demands that
the thermoelectric coefficient should be calculated in the next
approximation in the small parameters ${kT}/\zeta_{0}$,
${kT}/\Delta$, and ${kT}/{\mu^{\ast }B}$. However, it is not a
subject of this paper.

\section{Conclusions}

Hence, it is demonstrated that  the layered-structure effects at
closed Fermi surfaces and in weak quantizing magnetic fields can
manifest themselves as a phase delay of oscillations of the power
factor, an increase of the relative contribution of oscillations to
the power factor magnitude, and a simultaneous reduction of the
latter. However, in stronger magnetic fields, namely, at the point
where the chemical potential of the system of charge carriers has
the last oscillation maximum, the power factor reaches its maximum
value. The layered-structure effects in strong magnetic fields near
the ultraquantum limit reveal themselves as more pronounced
oscillations of the power factor, a reduction of the maximum value,
a smearing of the peak in the dependence of the power factor on the
magnetic field, and a slower recession of the power factor after
passing through the maximum.

In weak quantizing magnetic fields,  the power factor tends to zero
obeying the law $P\propto B^{3}$. In strong magnetic fields, near
the ultraquantum limit, it vanishes following the law $P\propto
T^{-3}B^{-6}$. At the same time, the amplitude of the power factor
maximum in perfect enough specimens with the ratio $l/a\geq30000$
can be comparable with the corresponding values for the best
cuprate-based thermoelectric materials \mbox{at 1000$~\mathrm{K}$.}

It is clear that all the results of  this paper need an experimental
verification. However, to the author's knowledge, all relevant
experiments concern galvano-magnetic effects in layered crystals
with a strongly open FS in the interval where the quasiclassical
approximation is applicable. One of a few exceptions is an old work
\cite{17} dealing with the Shubnikov--de Haas effect in graphite
intercalated with bromine, where just the topological transition was
considered from the open to the closed FS at the growth of the
bromine concentration.

It is rather difficult for the author  to judge about direct
practical applications of the results of his theoretical researches.
Nevertheless, he hopes that this research will stimulate the
statement of new experiments to study thermoelectric and
thermomagnetic phenomena in layered crystals, including those with a
closed FS. The latter circumstance, in the author's opinion, can be
important for the development of novel thermoelectric materials.

\vspace*{-5mm} \rezume{П.В.
Горський} {%
ФАКТОР ПОТУЖНОСТІ\\ ШАРУВАТОГО ТЕРМОЕЛЕКТРИЧНОГО\\ МАТЕРІАЛУ ІЗ
ЗАМКНЕНОЮ ПОВЕРХНЕЮ\\ ФЕРМІ В КВАНТУЮЧОМУ МАГНІТНОМУ
ПОЛІ}{Досліджено залежність фактора потужності шаруватого
термоелектричного матеріалу із замкненою поверхнею Фермі в
квантуючому магнітному полі при гелієвих температурах для випадку,
коли магнітне поле та градієнт температури спрямовані
перпендикулярно до шарів. Розрахунки проведено в наближенні сталого
часу релаксації. У слабких магнітних полях ефекти шаруватості
виражаються у відставанні осциляцій фактора потужності за фазою,
збільшенні їх відносного внеску і у деякому зменшенні величини
фактора потужності в цілому. В області сильних магнітних полів існує
оптимальний діапазон, в якому фактор потужності різко зростає і
досягає максимуму, абсолютна величина котрого при вибраних
параметрах задачі в наближенні ефективної маси на 12{\%} більша, ніж
для реального шаруватого кристала. Попри низькі температури, при цих
параметрах максимум фактора потужності в магнітному полі з індукцією
близько 1 Тл досягає значення, характерного для перспективних
купратних термоелектричних матеріалів при температурі понад 1000 К.
Для цього необхідно, щоб відношення довжини вільного пробігу носіїв
заряду до відстані між сусідніми шарами становило 30000 або більше.
Однак в ультраквантових магнітних полях фактор потужності різко
знижується за законом $P\propto T^{-3}B^{-6}$. Основною причиною
такого зниження є стиск поверхні Фермі в напрямку магнітного поля в
ультраквантовій границі внаслідок конденсації носіїв заряду на дні
єдиної заповненої підзони Ландау з номером $n=0$.}

\end{document}